\documentclass[12pt]{iopart}
\usepackage{iopams, graphics, epsfig}

\def\grg{{\it Gen. Rel. Gravit.}\ }

\def\be{\begin{equation}}
\def\ee{\end{equation}}

\begin{document}

\title{Averaging inhomogeneities in scalar-tensor cosmology}

\author{Vincenzo Vitagliano${}^1$, Stefano Liberati$^1$ and Valerio
Faraoni${}^2$}

\address{$^1$ SISSA - International School for Advanced Studies, Via
Beirut 2-4, 34151, Trieste, Italy and  INFN, Sezione di
Trieste\\
${}^2$Physics Department, Bishop's University, 2600 College St.,
Sherbrooke, Qu\'{e}bec, Canada J1M~1Z7}

\eads{\mailto{vitaglia@sissa.it}, \mailto{liberati@sissa.it},
\mailto{vfaraoni@ubishops.ca}}
\date{}

\begin{abstract}
The backreaction of inhomogeneities on the cosmic
dynamics is studied in the context of scalar-tensor gravity. Due
to terms of indefinite sign in the non-canonical effective energy
tensor of the Brans--Dicke-like scalar field, extra contributions
to the cosmic acceleration can arise. Brans--Dicke and metric $f(R)$ gravity are presented as specific examples. Certain representation
problems of the formalism peculiar to these theories are pointed
out.
\end{abstract}
\pacs{98.80.-k, 98.80.H, 04.20.-q, 04.90.+e}
%Cosmology
%Mathematical and relativistic cosmology
%Classical GR
%Other topics in GR and Gravitation

\section{Introduction}

The latest cosmological data sets and the increasing
number of ongoing  satellite missions dedicated to cosmology
 are poised to
raise a radically new theoretical scenario  as opposed to
the  description proper of the classical  General Relativity (GR)
schemes.  The cosmic acceleration detected by supernova
surveys \cite{SN} provides the starting point for a New Deal  in
cosmology, since dark energy and dark matter components seem to be needed in order to reproduce the observed phenomenology. Over the last
decade,  there have been many attempts to build models of
effective  fluids playing the role of dark energy: the taxonomy
of possible explanations includes the  resurrection of
Einstein's  cosmological constant (\cite{Linderresletter} and
reference therein), as well as the introduction of large-scale
modifications of gravity \cite{CCT,CDTT}.  Recently, a new
proposal  about the nature of the current cosmic
acceleration has been advanced, involving the backreaction of
inhomogeneities \cite{Buchert2, rasanen, Buchertall,
Buchertexactmodels}  as a possible source.

Even if the assumptions of
spatial homogeneity and isotropy  of the matter distribution
inspired by the Cosmological Principle
appear to give an adequate, although approximate,
description of  the universe on large scales, the
lumpiness of structures and the existence of huge voids  are
well-known observable properties  in smaller regions and at
late epochs. The
fitting problem, {\em i.e.}, the problem  of matching
 a coarse-grained matter distribution with a spacetime
metric obtained with an independent smoothing operator,
has been pointed out in  Refs.~\cite{ellis, ellis2}. The
development of an averaging procedure --- smoothing out
inhomogeneities of scalar
quantities --- allows us to implement a new set of averaged
contracted Einstein equations. The lack of commutativity between
time evolution and the averaging
procedure enables the encoding of  the kinematics
of the universe in terms of new quantities  with
recognizable backreaction features.

While the averaging formalism is interesting in itself,
and the idea of explaining the  cosmological data through
backreaction  in the context of pure Einstein gravity with no
dark energy  is very appealing, it has not been demonstrated yet
that this idea  works in practice. It is undeniable
that matter inhomogeneities  have a backreaction effect but it is
not clear that over/under-densities such as  those observed
around us are sufficiently large to significantly affect
the cosmic dynamics, and are not limited to small perturbative
effects. While the jury  is still out on whether backreaction
explains the observed   cosmic acceleration or not, one realizes
that virtually all high  energy theories attempting to quantize
gravity or unifying it  with the other interactions predict
deviations from GR. In string theories and
supergravity the gravitational field includes, in addition to
the massless spin two graviton, a dilaton whose presence is
unavoidable and that couples non-minimally to the curvature of
spacetime \cite{GreenSchwarzWitten}. Such a behaviour is mimicked
by scalar-tensor gravity \cite{BD, ST} (for example,  an
early representative of string theories, the bosonic string
theory reduces to an $\omega=-1$
Brans--Dicke theory in the low-energy limit \cite{bosonic}). 

While scalar-tensor theories
are constrained on Solar System scales and by the binary
pulsar \cite{Will}, we do not have many
constraints on larger scales (except, possibly, those due to the
variation of the effective gravitational coupling during Big Bang
nucleosynthesis). It is possible, therefore, that the
backreaction idea may have to be implemented in alternative
theories of gravity. In fact, it could even be that, if
backreaction doesn't quite work in  GR, it
is ``helped'' by a non-Einsteinian component of  gravity. In
\cite{CardoneEsposito} a  formalism that
implements Buchert's scheme into models with variable  Newton
``constant'' was already developed, motivated by the
non-perturbative  renormalization group improvement of the action
functional \cite{ren-group}. Here, instead, we restrict our attention
to   scalar-tensor gravity as the prototypical generalization of
GR.

The following observation can be made  {\em a priori}: the
Brans--Dicke-like field that necessarily permeates all of
spacetime can be described  as an effective form of matter by
writing the scalar-tensor field  equations in the form of
effective Einstein equations.  The effective energy-momentum
tensor characterizing this form of  $\phi$-matter 
easily violates all the energy conditions  and, therefore, is
more likely to produce the
cosmic acceleration.

Another aspect is worth pointing out: it is widely
believed that quantum corrections to the Einstein--Hilbert action
introduce quadratic deviations from the usual Lagrangian density
$R$, which may well have propelled the inflationary epoch in the
early   universe,~\footnote[1]{We do not refer here specifically to $f(R)$
theories based on large-scale modifications of gravity \cite{CCT,
CDTT}. It would be rather pointless to study the
backreaction  effect in those
$f(R)$ theories since it is already known that, in their metric
version,  they  may provide viable models to explain the
cosmic acceleration~\cite{review}.} as in Starobinsky's
inflation \cite{Starobinsky}. For  a spatially homogeneous and
isotropic  universe, quadratic corrections die off quickly as the
universe expands and $R$ decreases. However,  in an inhomogeneous
universe,  they might help the
backreaction mechanism. Now, it is well-known
\cite{STequivalence, review}
that a theory described by a  non-linear Lagrangian density
$f(R)$ in the metric formalism is equivalent to an $\omega=0$
Brans--Dicke theory with a scalar field degree of freedom given by
$\phi=f'(R)$ with a suitable scalar field potential. Therefore,
by studying scalar-tensor theory, we also catch the
effect of the simplest quadratic corrections to GR.

The  scalar-tensor action  expressed in the
Jordan frame is
\be\label{BDaction}
S_{ST}=\int d^4x \sqrt{-g} \left\{\frac{1}{16\pi} \left[ \phi
R-\frac{\omega(\phi)}{\phi} \,
\nabla^{\alpha} \phi \nabla_{\alpha} \phi -V(\phi) \right]
+\alpha_m\mathcal{L}^m\right\} \,,
\ee
where $\phi$ is the Brans--Dicke-like scalar field with potential
$V(\phi)$ and coupling function $\omega(\phi)$, $g$ is the determinant of the metric tensor
$g_{\mu\nu}$, $R$ is the Ricci curvature, $\mathcal{L}^m$
is the Lagrangian density describing the ordinary matter sector with coupling costant $\alpha_m$, and we adopt the notations of
Ref.~\cite{Wald}.

The conformal transformation
\be
g_{\mu\nu}\rightarrow \tilde{g}_{\mu\nu}=\Omega^2 \, g_{\mu\nu}
\,, \;\;\;\;\;\;\;\; \Omega=\sqrt{G\phi}
\ee
and the scalar field redefinition
\be
 d\tilde{\phi}= \sqrt{
\frac{2\omega(\phi)+3}{16\pi G}} \, \frac{d\phi}{\phi}
\ee
turn the action~(\ref{BDaction}) into its Einstein
frame form
\be
S_{ST}=\int d^4x \,  \sqrt{-\tilde{g}} \left[ \frac{
\tilde{R} }{16\pi G} -\frac{1}{2} \, \tilde{g}^{\mu\nu}
\tilde{\nabla}_{\mu}\tilde{\phi} \,
\tilde{\nabla}_{\nu}\tilde{\phi} \,-U\left( \tilde{\phi} \right)
+\tilde{\alpha}_m(\tilde{\phi})\mathcal{L}^m\right]\,,
\ee
where a tilde denotes quantities in the rescaled world, and
\be
U\left(\tilde{\phi} \right)=\frac{ V[\phi(\tilde{\phi})] }{\left[
G\phi(\tilde{\phi}) \right]^2} \,,\;\;\;\;\;\;\;\;\tilde{\alpha}_m(\tilde{\phi})=\frac{\alpha_m}{\left[
G\phi(\tilde{\phi}) \right]^2}\,.
\ee
The ``new'' scalar field $\tilde{\phi}$ couples minimally to the
curvature but non-minimally to the  matter fields.

\section{Averaging procedure for GR cosmology}

Our goal is studying the backreaction mechanism of spatial inhomogeneities on the cosmic dynamics in the context of scalar-tensor gravity. Before doing this, we briefly review the Buchert formalism in GR for a universe filled with an irrotational dust. In this case it is
possible to  choose a foliation of spacetime  with
spacelike hypersurfaces orthogonal to the flow at any event. We
will then apply  the averaging procedure with respect to a family
of observers comoving with the dust and characterized by a
four-velocity field $u^{\mu}$, thus avoiding gauge complications
related to the choice of an arbitrary set of  observers
tilted
with respect to the cosmological matter fluid \cite{gauge}.
In actual fact, in an inhomogeneous universe the
four-velocity of these observers is not simply
$ u^{\mu}=\delta^{0\mu}$ but there are also local fluctuations
$\delta u^{\mu}$, so that $u^{\mu}=\delta^{0\mu}+\delta u^{\mu}$
corresponding to the possible choices of time on the
inhomogeneous  hypersurfaces. Therefore, the procedure adopted here of projecting the
Einstein equations onto $u^{\mu}$ and then averaging is not free
of ambiguities and gauge-dependence issues. This projection and
the spatial average do not commute. With this caveat in mind, we proceed as is usually
done in the literature by choosing Gaussian normal coordinates
(see below).

It is also convenient to define a template metric
mimicking  the main properties of a FLRW universe
on large scales \cite{Larena08, para} but encoding
the small scale lumpy structures.  In this way the
averaged quantities will assume the
usual meaning as in the traditional cosmological framework.  The
scale  of the domain used in the averaging
procedure is chosen as the cosmological volume over which
it would be reasonable to recover homogeneity,
{\em  i.e.}, somehow larger than  $100 ~h^{-1}$~Mpc.

Let us briefly recall the essential points of Buchert's averaging
approach, referring the reader to \cite{Buchert1} for details.
For the sake of simplicity we turn our attention to
Buchert's original model (see \cite{Buchert2} for a comprehensive
review). This  consists of a  universe filled with an
irrotational dust as the  material  source, with energy
density  $\rho$ and four-velocity $u^{\mu}$ satisfying $
u_{\mu}u^{\mu}=-1$. The corresponding Einstein
equations and stress-energy covariant conservation equation read
\begin{eqnarray}
&& R_{\mu\nu}-\frac{1}{2} \, g_{\mu\nu}R=8\pi G
\, \rho \, u_{\mu}u_{\nu}-\Lambda g_{\mu\nu}\; ,\label{11}\\
&& \nonumber \\
&& \nabla_{\mu}\left(\rho \, u^{\mu}u^{\nu}\right)=0
\,,\label{a8}
\end{eqnarray}
where $\rho \equiv T_{\mu\nu}u^{\mu} u^{\nu}$. By adopting
Gaussian  normal coordinates it is possible to apply  the
standard  ADM procedure for the 3+1 splitting of spacetime
\cite{Wald}. In these  coordinates the spacetime manifold can be
foliated with spacelike Cauchy hypersurfaces parametrized by the
proper time $t$. In this framework the surfaces are comoving
with the fluid in such a way that, casting the metric in the
form
\begin{equation}
ds^2=-dt^2+g_{ij}\left(t, X^k\right) dX^i\otimes
dX^j    \hspace{3cm} (i,j,k=1,2,3),\label{pinco1}
\end{equation}
we have $u^{\mu}=\left( 1, 0, 0, 0 \right)$ and $u^{\nu}
\nabla_{\nu} u^{\mu}=0$. The second fundamental form (extrinsic
curvature) $K_{\mu\nu}$ of the geodesic normal slicing of
spacetime is introduced as follows: Let
$h_{\mu\nu}=g_{\mu\nu}+u_{\mu} u_{\nu}$ be the induced metric on
the  3-surfaces. Then $K_{\mu\nu}$ is defined as the Lie
derivative of  this Riemannian metric in the time direction,
\begin{equation}
K_{\mu\nu}=
-\frac{1}{2}\pounds_uh_{\mu\nu}=-\nabla_{\mu}u_{\nu}
=-\frac{1}{2}\partial_th_{\mu\nu}\,. \label{pinco3}
\end{equation}
Given the form of the metric (\ref{pinco1}), $K_{00}$ and
$K_{0i}$ vanish while $K_{ij}$ can be    expressed in terms of
the expansion tensor $\theta_{ij}$, the expansion scalar $\theta \equiv {\theta^i}_i$, and the traceless shear tensor $\sigma_{ij}$ as
\be\label{a4}
K_{ij} = -\theta_{ij}= -\left(\sigma_{ij}+\frac{ \theta}{3}  \,
g_{ij}\right)\,, \;\;\; K\equiv K_i^{\phantom{i}i} =- \, \theta
\;\;\;\;\; (i,j=1,2,3).
\ee
Denoting with $D_{\mu}$ the derivative operator associated with the
metric $h_{\mu\nu}$, it is possible to derive the Gauss--Codazzi
relations between the  curvature of the 3-surface, the extrinsic
curvature and the spacetime curvature \cite{Wald}:
\be
{}^{(3)}R_{\mu\nu\rho\sigma}=
{}^{(4)}R_{\alpha\beta\gamma\delta}
h^{\alpha}_{\phantom{\alpha}\mu} h^{\beta}_{\phantom{\beta}\nu}
h^{\gamma}_{\phantom{\gamma}\rho}
h^{\delta}_{\phantom{\delta}\sigma}  -K_{\mu\rho}
K_{\nu\sigma}+K_{\mu\sigma} K_{\nu\rho} \,, \label{gauss1}
\ee

\be
D_{\rho}K^{\rho}_{\phantom{\rho}\nu}-
D_{\nu}K=h^{\mu}_{\phantom{\mu}\nu}R_{\mu\rho}u^{\rho} \,.
\ee
Saturating indices  with the induced metric $h_{\mu\nu}$, it is
possible to rearrange eq.~(\ref{gauss1}) as
\be
G_{\mu\nu} u^{\mu}u^{\nu} =
\frac{1}{2}\left(^{(3)}\mathcal{R}+K^2
-K_{ij}K^{ij}\right) \,, \label{gauss2}
\ee
where $^{(3)}\mathcal{R}$  is the scalar 3-curvature, {\em i.e.},
the projection of the Ricci  scalar onto the spatial
hypersurface. On the other hand, using the definition of
the
Riemann tensor it follows that
\be
R_{\mu\nu}u^{\mu}u^{\nu}=K^2-K_{\mu\nu}K^{\mu\nu}-\nabla_{\mu}\left(u^{\mu}\nabla_{\nu}u^{\nu}\right)
+\nabla_{\nu}\left(u^{\mu}\nabla_{\mu}u^{\nu}\right)\,,
\label{pinco2}
\ee
with the last term vanishing because of the geodesic equation
obeyed by the four-velocity of the dust.
By combining~(\ref{pinco2}) with~(\ref{gauss2}) and taking
into  account the definition~(\ref{pinco3}) of extrinsic
curvature,
we are able to express the scalar curvature  of spacetime as
\be
{}^{(4)}R={}^{(3)}\mathcal{R}+K^2+K_{ij}K^{ij}-2\pounds_uK\,.
\label{gauss3}
\ee
The Hamiltonian or energy  constraint and the evolution equation
for the expansion scalar (Raychaudhuri equation) can be
derived from
appropriate contractions of the Einstein equations:
the Hamiltonian constraint is obtained by doubly 
contracting eq.~(\ref{11}) with $u^{\mu} $ and using
eq.~(\ref{gauss2}),
\begin{equation}\label{a2}
\frac{1}{2}\left(^{(3)}\mathcal{R}+K^2
-K_{ij}K^{ij}\right)=8\pi G \rho+\Lambda\, ,
\end{equation}
while the equation for the scalar expansion is found
by tracing
the Einstein  equation.
Taking into account eq.~(\ref{gauss3}) and the fact that
$\pounds_uK=\partial_tK$, it follows  that
\begin{equation}\label{a3}
^{(3)}\mathcal{R}+K^2+K_{ij}K^{ij}-2\partial_tK=8
\pi G\rho+4\Lambda\, .
\end{equation}
The scheme proposed by Buchert involves scalar
quantities averaged over a compact  domain $D$ with volume
$V_D\equiv\int_D d^3X \, \sqrt{^{(3)}g}  $,

\begin{equation}\label{a12}
\left\langle \psi(t, X_i)\right\rangle_D \equiv
\frac{1}{V_D}\int_D d^3X \,\sqrt{^{(3)}g}\,  \psi \left(t,X_i
\right)  \,.
\end{equation}
Hence, in order to apply the averaging procedure, it is
useful to re-arrange eqs.~(\ref{a2}) and (\ref{a3}) taking into
account the relations~(\ref{a4}). In this way, we
find
the scalar equations~\footnote[2]{Hereafter
an overdot denotes differentiation   with respect
to the  comoving time $t$ and the Latin indices
$i$ and $j$ assume the values ~1, 2, and~3.}
\begin{eqnarray}
\frac{1}{2}\left(^{(3)}\mathcal{R}+
\frac{2}{3} \, \theta^2-2\sigma^2\right)
=8\pi G\rho+\Lambda  \,, \label{a6}\\
\nonumber \\
{}^{(3)}\mathcal{R}
+\frac{4}{3} \, \theta^2+2\sigma^2+2\dot{\theta}
=8\pi G\rho+ 4\Lambda\,,\label{a7}
\end{eqnarray}
where we have defined the  shear scalar as
$\sigma^2\equiv\frac{1}{2}
\sigma_{ij}\sigma^{ij}$.  It is also
useful to recall the energy conservation equation~(\ref{a8}),
which takes the form
\be\label{a9}
\dot{\rho}=K\rho=-\theta\rho \,.
\ee
In a spatially homogeneous and isotropic  universe with
curvature  index $\kappa$  described by the
Friedmann-Lemaitre-Robertson-Walker (FLRW) metric 
\footnote{The 
Buchert scheme applies to vorticity-free spacetimes and it is not 
clear how to fit a small amount of rotation into a Buchert-like 
scheme. This issue deserves some attention in the future.} 
\be ds^2=-dt^2+a^2(t) \left[ \frac{dr^2}{1-\kappa r^2}+r^2 \left(
d\theta^2+\sin^2\theta d\varphi^2 \right) \right]
\ee
and dominated by dust, one has
\begin{eqnarray}
&& \left( \frac{\dot{a}}{a} \right)^2=\frac{8\pi G
\rho}{3}+\frac{\Lambda}{3} -\, \frac{\kappa}{a^2} \,,
\label{B22-1}\\
&&\nonumber\\
&& \frac{\ddot{a}}{a}=-\, \frac{4\pi}{3} G\rho
+\frac{\Lambda}{3}
\,, \label{B22-2}\\
&&\nonumber\\
&& \dot{\rho}+3 \, \frac{\dot{a}}{a}\, \rho=0 \,. \label{B22-3}
\end{eqnarray}
Using the averaging procedure, eqs.~(\ref{a6})-(\ref{a9})
can always be written in the form of a
Friedmann-like system of averaged equations, following the
operational definition~(\ref{a12}) and exploiting the
non-trivial commutation relation that holds for any scalar
quantity $\psi(t, X_i)$ \cite{Buchert1}
\begin{equation}
\langle\psi(t, X_i)\rangle^{\cdot}_D-\langle\dot{\psi}(t,
X_i)\rangle_D =\langle\psi(t, X_i)\theta\rangle_D-\langle\psi(t,
X_i)\rangle_D\langle\theta\rangle_D\,.
\end{equation}
Let us introduce also a dimensionless scale factor normalized by
the volume $V_{D_i}$ of the
region $D$ at some initial time $t_i$ as
$ a_D(t)\equiv\left(V_D/V_{D_i}\right)^{1/3}$, with the
property that  the  averaged expansion rate is written
as
\be
\langle\theta\rangle_D=
\frac{\dot{V}_D}{V_D} = 3 \, \frac{ \dot{a}_D}{ a_D}
\equiv3H_D \,.
\ee
We define a ``kinematical backreaction'' term, vanishing on a
FLRW background, as
\be\label{a13}
\mathcal{Q} _D\equiv\frac{2}{3} \left(\langle
\theta^2\rangle_D-\langle\theta\rangle_D^2\right)
-2\langle\sigma^2\rangle_D = \frac{2}{3}\langle \theta^2
\rangle_D -2\langle \sigma^2 \rangle_D -6H_D^2 \,.
\ee
The Einstein scalar equations and the covariant conservation
equation now yield
\begin{equation}\label{a10}
3\left(\frac{\dot{a}_D}{a_D}\right)^2-8\pi
G\left\langle\rho\right\rangle_D-\Lambda=
-\frac{\left\langle^{(3)}\mathcal{R}\right\rangle_D+\mathcal{Q}_D}{2}\,,
\end{equation}
\begin{equation}\label{a11}
3 \, \frac{\ddot{a}_D}{a_D}+4\pi
G\left\langle\rho\right\rangle_D-\Lambda=\mathcal{Q}_D\,,
\end{equation}
\be\label{a15}
\left\langle\dot{\rho}\right\rangle_D
+\left\langle\theta\rho\right\rangle_D=
\left\langle\rho\right\rangle_D^{\cdot}+3
\, \frac{\dot{a}_D}{a_D}\left\langle\rho\right\rangle_D=0 \,,
\ee
respectively.
The energy constraint (\ref{a10}) and the Friedmann
acceleration law (\ref{a11})
lead to a  differential integrability condition
involving $ \mathcal{Q}_D$ and  $
\left\langle^{(3)}\mathcal{R}\right\rangle_D$ that
accounts for the coupling between 3-curvature and fluctuations:
\begin{equation}\label{a14}
\frac{1}{a_D^6} \, \partial_t \left(\mathcal{Q}_D \,a_D^6\right)+
\frac{1}{a_D^2} \,
\partial_t\left(\left\langle^{(3)}\mathcal{R}\right\rangle_D\,
a_D^2\right)=0\,.
\end{equation}
The system of averaged equations is not closed because there
are only three independent equations for the four
unknown functions $a_D, \left\langle\rho\right\rangle_D,
\mathcal{Q}_D, \left\langle^{(3)}\mathcal{R}\right\rangle_D$.
This means that, in principle, different spacetime{s} could
evolve  in  different ways even when they have the same
average  initial conditions.  Extra  assumptions are
needed to close the system,  for  example assuming a certain
effective cosmic equation of state, or demanding
a particular functional relationship between $\mathcal{Q}_D$
and  $ \left\langle^{(3)}\mathcal{R}\right\rangle_D$ (as it is
done in \cite{Buchert1,BuchertLarenaAlimi06} in order to obtain scaling solutions).

\section{Averaging procedure for scalar-tensor
cosmology}

It is convenient to write the field equations of scalar-tensor
gravity in the form of effective Einstein equations, which allows
for the direct application of Buchert's formalism to this  class
of  theories. It must be pointed out that choosing this form of
the equations  implies that the scalar field
$\phi$ plays the role of the inverse of a
Newton
``constant'' now varying in space and time  (the effective
gravitational coupling in the
action~(\ref{BDaction}) is $G_{eff}=\phi^{-1} $, although the
coupling in a Cavendish experiment is instead
$G_{eff}=\frac{1}{\phi}\, \frac{ 2( \omega+2)}{2\omega+3} $
\cite{coupling}).
It is rather simple to notice that the presence of this extra field introduces a new ambiguity with respect to GR due to the non-linearity of the averaging procedure.
In fact, the variation of the action~(\ref{BDaction}) with
respect to
$g^{\mu\nu}$ yields the field equations
\be\label{Delta}
 \phi G_{\mu\nu} =  8\pi \left( T_{\mu\nu}^{(m)}+
T_{\mu\nu}^{(\phi)} \right) \,,
\ee
where $G_{\mu\nu}\equiv R_{\mu\nu}-\frac{1}{2}\, g_{\mu\nu}R$ is
the Einstein tensor and
\be
\hspace{-0.5cm}T_{\mu\nu}^{(\phi)} =
\frac{\omega(\phi)}{\phi}\left(\nabla_{\mu}\phi\nabla_{\nu}\phi
-\frac{1}{2}g_{\mu\nu}\nabla^{\sigma}\phi\nabla_{\sigma}\phi\right)
+ \nabla_{\mu}\nabla_{\nu}
\phi-g_{\mu\nu}\Box
\phi  -\frac{V(\phi)}{2} \, g_{\mu\nu}\; .
\ee
While it is common to divide by $\phi$ to put this equation in
the form of the effective Einstein equation
\begin{eqnarray}\label{4}
&& R_{\mu\nu}-\frac{1}{2}g_{\mu\nu}R=
\frac{8\pi}{\phi} \, T_{\mu\nu}^{(m)}+
\frac{\omega(\phi)}{\phi^2}\left(\nabla_{\mu}\phi\nabla_{\nu}\phi
-\frac{1}{2}g_{\mu\nu}\nabla^{\sigma}\phi\nabla_{\sigma}\phi\right)
\nonumber\\
&& \hspace{5.5cm}+\frac{1}{\phi}\left(\nabla_{\mu}\nabla_{\nu}
\phi-g_{\mu\nu}\Box
\phi\right) -\frac{V(\phi)}{2\phi} \, g_{\mu\nu}\; ,
\end{eqnarray}
this operation does not commute with the spatial average if
$\partial\phi/\partial x^i \neq 0$. As a result, once the scalar averaging has been performed,
$ \langle \phi\; {}^{(4)}R \rangle_D \neq
\langle \phi \rangle_D  \langle  {}^{(4)}R \rangle_D $. This
problem does not appear in GR where the coupling
is a true constant and is peculiar to scalar-tensor gravity.
The outcomes of taking  the average of eq.~(\ref{Delta}) or of
eq.~(\ref{4}) are different. For ease of comparison with GR we choose to proceed by averaging eq.~(\ref{4}) but
with a second caveat to keep in mind. Further, if one decides
to adopt the Einstein conformal frame instead of the Jordan
frame, the relevant  integro-differential equations  can,
in principle, have  different solutions in the two frames.
But this ambiguity remains even if we stay in the Jordan frame,
depending on the choice one makes to use the scalar field
directly  linked to the gravitational sector  or, as in our case,
to recast the field equations as effective Einstein-like
equations.

The variation of  the action~(\ref{BDaction})
with respect to  the scalar field yields the equation of
motion for $\phi$
\be \label{BBBox}
\Box\phi=\frac{1}{2\omega(\phi)+3}\left[ -8\pi \rho-\frac{d\omega}{d\phi}\nabla^{\sigma}\phi\;\nabla_{\sigma}\phi+\phi \, \frac{d
V(\phi)}{d\phi}-2V(\phi)\right] \,.
\ee
The Hamiltonian constraint is obtained by double contraction of
the previous equation with $u^{\mu} $ (time-time component
of the field equations)
\begin{eqnarray}\label{5}
&& \frac{1}{2}\left(^{(3)}\mathcal{R}
+K^2-K_{ij}K^{ij}\right)=
\frac{8\pi\rho}{\phi}+\frac{\omega(\phi)}{2} \,\frac{
\dot{\phi}^2}{\phi^2} +
\frac{\omega(\phi)}{2\phi^2} \, g^{ij} \partial_i \phi\partial_j \phi
\nonumber\\
&&  \hspace{7.5cm}+\frac{1}{\phi}
\left(\ddot{\phi}+\Box\phi\right)+\frac{V(\phi)}{2\phi} \,,
\end{eqnarray}
while the evolution equation for the expansion scalar now reads
\begin{eqnarray}\label{8}
&& ^{(4)}R=^{(3)}\mathcal{R}+K^2+K_{ij}K^{ij}-2\partial_t K
\nonumber\\&& \nonumber\\
&& =-g^{\mu\nu}\left[\frac{8\pi}{\phi}T_{\mu\nu}^{(m)}+
\frac{\omega(\phi)}{\phi^2}\left(\nabla_{\mu}\phi\nabla_{\nu}\phi
-\frac{1}{2}g_{\mu\nu}\nabla^{\sigma}
\phi\nabla_{\sigma}\phi\right)\right. \nonumber\\&& \nonumber\\
&& \hspace{4cm} +\left.\frac{1}{\phi}
\left(\nabla_{\mu}\nabla_{\nu}\phi-g_{\mu\nu}\Box
\phi\right)-\frac{V(\phi)}{2\phi}g_{\mu\nu} \right]
\nonumber\\&& \nonumber\\
&& =8\pi\frac{\rho}{\phi}+\frac{\omega(\phi)}{\phi^2}\nabla^{\mu}
\phi\nabla_{\mu}\phi+
\frac{3\Box\phi}{\phi} +\frac{2V(\phi)}{\phi}\,.
\end{eqnarray}
By averaging the last two equations and using both the
definition~(\ref{a13})  of backreaction  and
the fact  that $ K^2-K_{ij}K^{ij}=\frac{2}{3} \, \theta^2-
2\sigma^2$, one obtains
\begin{eqnarray}
&& \hspace{-1cm}\frac{1}{2}\left
\langle^{(3)}\mathcal{R}\right\rangle_D
+\frac{1}{2}\mathcal{Q}_D +3H_D^2=
8\pi\left\langle\frac{\rho}{\phi}
\right\rangle_{\!\!D}+
\left\langle\frac{\omega(\phi)}{2}\;\frac{\dot{\phi}^2+
g^{ij}\partial_i\phi\partial_j\phi}{\phi^2}
\right\rangle_{\!\!D} \nonumber\\&& \nonumber\\
&& \hspace{5cm}+\left\langle\frac{\ddot{\phi}
+\Box\phi}{\phi} +\frac{V(\phi)}{2\phi}\right\rangle_{\!\!D}\,,
\label{6}\\
&& \nonumber\\
&& \hspace{-1cm}\left\langle^{(3)}\mathcal{R}\right\rangle_D-
\mathcal{Q}_D +6H_D^2+6 \, \frac{\ddot{a}_D}{a_D}
=8\pi\left\langle\frac{\rho}{\phi}\right\rangle_{\!\!D}+
\left\langle\omega(\phi)\left(\frac{-\dot{\phi}^2+
g^{ij}\partial_i\phi\partial_j\phi}{\phi^2}\right)\right\rangle_{\!\!D}
\nonumber\\
&& \nonumber\\
&& \hspace{5.5cm}+\left\langle
\frac{3\Box\phi+2V(\phi)}{\phi}\right\rangle_{\!\!D}\label{12}\,.
\end{eqnarray}
By combining the last two equations and using
eq.~(\ref{BBBox}) the cosmic acceleration
is  expressed as
\begin{eqnarray} \label{mammamia}
\hspace{-2cm}\frac{ \ddot{a}_D }{a_D}= - \frac{8\pi}{3}
\left\langle \frac{\rho}{\phi}\left(\frac{\omega(\phi)+2}{2\omega(\phi)+3}\right)
\right\rangle_D
+ \frac{ \mathcal{Q}_D }{3}
-\frac{1}{3} \left\langle\omega(\phi)
\left( \frac{\dot{\phi}}{\phi} \right)^2 \right\rangle_D
- \frac{1}{3} \left\langle \frac{ \ddot{\phi} }{\phi } \right\rangle_D\\&&\nonumber\\
\hspace{-1.5cm}-\frac{1}{6}\left\langle \frac{1}{2\omega(\phi)+3}\,\frac{d\omega}{d\phi}\,\nabla^{\sigma}\phi\;\nabla_{\sigma}\phi
\right\rangle_D +\frac{1}{6}\left\langle \frac{1}{2\omega(\phi)+3}\left( \frac{dV}{d\phi}+\left(2\omega(\phi)+1\right)\frac{V}{\phi}\right)
\right\rangle_D\,. \nonumber
\end{eqnarray}
Since $\phi>0$ and $\omega(\phi)>0$ in order to keep the gravitational
coupling positive, the positive energy density of dust in the 
first  term on the right hand side causes deceleration. 

The constraints on the magnitude of the factor  ${ 2(\omega+2)}/{(2\omega+3)}$ depend on the range of the $\phi$.
If the latter is comparable with the size of the solar system then the Cassini bound $\omega>40000$ \cite{BertottiIessTortora} applies. 
However, this bound does not apply if the field is short-ranged or if endowed with a range depending on the environment (chameleon mechanism).

In an optimistic view, the backreaction term 
$\mathcal{Q}_D$ is  positive
and contributes to acceleration, as generally argued in GR. 
However, this is not necessarily the case: in fact, prior to the 
1998 discovery of the cosmic acceleration, the same backreaction 
term, with negative sign, was proposed as a solution to the dark 
matter problem (see \cite{BuchertDM} and Sec.~5.5.2 of 
\cite{Tsagasetal08}).  This shows that the sign of ${\cal Q}_D$ 
is highly uncertain. The third term on the right hand side of
eq.~(\ref{mammamia}) is definitely negative and
contributes to decelerate the universe, while the signs of
the fourth and fifth terms are undetermined. 

There is little doubt that the terms involving the
first and second derivatives of $\phi$ are small and, at best
({\em i.e.}, when $ \langle \ddot{\phi} \rangle_D<0$)
their effects conflict. However, the constraints on the temporal
and spatial variation of $\phi$ after nucleosynthesis are
rather poor. While the time variation of the gravitational
coupling is constrained as $ \left| \frac{\dot{G} }{G}  \right|
\simeq \left| \frac{\dot{\phi}}{\phi} \right|
< H_0^{-1}$ (where $H_0$ is the present value of the Hubble
parameter)~\cite{Will}, there is basically no constraint on the second time derivative of $\phi$. 

The last term including the potential and its derivative is novel with respect to GR and could significantly affect the acceleration. While this could be interpreted as an obvious consequence of the fact that a potential can mimic a cosmological constant, we show later (see the case of $f(R)$ gravity discussed below) that it can be important and positive even in cases for which late time acceleration cannot be {\it a priori} expected from the form of the Lagrangian.

%could be regarded as an effective cosmological constant
%which is slowly varying since $\phi$ is not constant (but its variations at
%the present epoch are constrained to be small). 
%Retaining this
%term would amount to introduce a cosmological constant in the
%field equations, but this can accelerate the expansion of the
%universe without backreaction or scalar-tensor gravity, so it is
%rather pointless to include this term, which will be dropped from
%now on.
In summary,  while no definitive conclusion can be reached on whether the inclusion of backreaction induces late time acceleration (as in the GR case), nonetheless there are encouraging new terms in scalar-tensor cosmology.
Unfortunately no definitive answer on the relative magnitude and sign of the specific terms can be provided in such a general framework.
Hence, in the following we shall consider specific implementation of the theory in which eq.~(\ref{mammamia}) simplifies.

\subsection{Brans--Dicke cosmology}

As an example of the procedure developed, let us specialize 
the whole formalism to a true Brans--Dicke theory ({\em i.e.}, 
$ V \equiv 0$ and $\omega(\phi)\equiv\omega_0=\textrm{constant}$) and let us also assume the scalar field 
to be spatially smooth on the scales of
interest,  $ \phi=\phi(t)$. This is clearly an
oversimplification  but
serves the purpose of illustration. This
assumption implies that all the averages involving the
scalar field $\phi$ are domain-independent. In this context, the
ambiguity in the choice of the representation  described
in  the previous section is no longer present.  Then,
eqs.~(\ref{6}) and (\ref{12}) become
\begin{eqnarray}\label{7}
&& \frac{1}{2}\left\langle^{(3)}\mathcal{R}
\right\rangle_D+\frac{1}{2}\mathcal{Q}_D+3H_D^2=
8\pi\frac{\left\langle\rho\right\rangle_D}{\phi}
+\frac{\omega_0}{2}
\left(\frac{\dot{\phi}}{\phi}\right)^2
-3H_D\frac{\dot{\phi}}{\phi} \,, \end{eqnarray}
\begin{equation}\label{13}
\frac{6\ddot{a}_D}{a_D}= - \left\langle^{(3)}\mathcal{R}
\right\rangle_D  + \mathcal{Q}_D - 6H_D^2
+ 8\pi \frac{\left\langle\rho \right\rangle_D}{\phi}
-\omega_0 \, \frac{\dot{\phi}^2}{\phi^2}-
3 \, \frac{ \left( \ddot{\phi}+3H_D\dot{\phi} \right)}{\phi} \,.
\end{equation}
The consistency relation between  the Hamiltonian constraint and
the Raychaudhuri equation can now be derived by differentiating
the latter with  respect to time and then substituting the
result, the Hamiltonian constraint,  and the equation of motion
for  the scalar field in the former. The result is
\begin{eqnarray}
&& \frac{1}{a_D^6} \,
\partial_t\left(\mathcal{Q}_D \,a_D^6\right)+
\frac{1}{a_D^2} \, \partial_t\left(\left\langle^{(3)}\mathcal{R}
\right\rangle_D\, a_D^2\right)= \\
%&&\nonumber \\
&& =\frac{2}{a_D^{\frac{6\omega_0+12}{2\omega_0+3}}} \, \partial_t
\left[8\pi\frac{\left\langle\rho
\right\rangle_D}{\phi}
\, a_D^{\frac{6\omega_0+12}{2\omega_0+3}}\right]+
\frac{1}{a_D^6} \, \partial_t\left[\frac{\omega_0
\, \dot{\phi}^2}{\phi^2} \,
a_D^6\right]-
\frac{6}{a_D^4} \, \partial_t\left[\frac{\dot{\phi}}{\phi}\,
H_D \, a_D^4\right] \,. \label{minchia}
\nonumber
\end{eqnarray}
As a check, it is noted that this equation reduces to the
corresponding eq.~(\ref{a14}) in the limit $\omega_0\rightarrow
\infty, \, \phi \approx \mbox{const.}+\mbox{O} \left(
\frac{1}{\omega_0}  \right) $ in which Brans--Dicke
theory reduces to GR~\footnote[3]{In the case of
a massive dust, the limit of  Brans--Dicke theory to GR is free of the  ambiguities arising when $T^{(m)}=0$
and the expansion $\phi=\mbox{const.}+\mbox{O}\left(
\frac{1}{\omega_0} \right)$ is indeed correct
(see \cite{BDlimitproblems} and  references therein).} (this can  be
seen by using the form of  the solution of
eq.~(\ref{a15}),
$\left\langle\rho\right\rangle_D \propto
a_D^{-3}$, in the first term on the right hand side of
eq.~(\ref{minchia})).

Let us consider a class of solutions in which the scalar field has the form
\be
\phi(t)=\phi_0+\phi_1 \mbox{e}^{-\beta t} \,,
\ee
where the requirement of a positive, non-vanishing scalar field implies $\phi_0\,,\beta>0$ and $\phi_1>-\phi_0$. Using the general solution of  eq.~(\ref{a15})
we   can express the  averaged energy density as
$\langle\rho\rangle_D(t) =\langle\rho\rangle_D^0 \, a_D^{-3}(t)$, where the scale factor has been normalized at the starting time of the growth of structures (in our notation, $a_D(t=0)=1$ where $t=0$ corresponds to the last scattering surface).
Inserting this relationship into the equation of motion for
$\phi$, it is
possible to solve with respect to $a(t)$. The effective gravitational
coupling is finite for both small and large times $t$, and the
corresponding averaged scale factor is
\be
a_D(t)=\mbox{e}^{\frac{\beta t}{3} } \left( 1-\gamma t \right)^{1/3}
\ee
with
\be
\gamma=\frac{ 8\pi \langle \rho \rangle^0_D}{\beta\phi_1
(2\omega+3)}\,.
\ee
It is an easy task to show that late time accelerated solutions can be found for suitable values of the parameters. However, the physically motivated requirement that the backreaction is negligible at early stages further restricts the allowed range.\footnote[4]{An example of such a solution can be found for the set of values  $(\beta, \phi_0, \phi_1, \omega, \langle\rho\rangle_D^0)=(0.002, 750,-1,40000,1)$.} 

The following expressions for $\langle\mathcal{R}\rangle_D$ and $\mathcal{Q}_D$ are immediately obtained:
\be
\hspace{-2cm}\langle\mathcal{R}\rangle_D=\frac{\beta \phi_1 \gamma-24\pi \langle \rho \rangle^0_D-2\beta e^{\beta t}\phi_0[\gamma(2+\beta t)-\beta]}{2\left(\phi_1+e^{\beta t}\phi_0\right)\left(\gamma t-1\right)}\,,
\ee
\begin{eqnarray}
\hspace{-2cm}\mathcal{Q}_D=\frac{\beta^2 \phi_1^2\omega}{\phi_1+e^{\beta t}\phi_0}+ \frac{-8\pi\langle \rho \rangle^0_D+\beta\phi_1[\gamma(2\beta t-1)-2\beta]}{2\left(\phi_1+e^{\beta t}\phi_0\right)\left(\gamma t-1\right)}+\nonumber\\
\hspace{5.5cm}+\frac{1}{3} \left[\beta^2+\frac{2\beta\gamma}{\gamma t-1}-\frac{2\gamma^2}{(\gamma t-1)^2}\right]\,.
\end{eqnarray}

The initial value of the backreaction term $\mathcal{Q}_D$ could be different from zero (albeit small), as long as we assume a perturbed FLRW universe at the last scattering epoch. Furthermore, $\mathcal{Q}_D$ approaches the asymptotic value ${\beta^2}/{3}$, giving a positive contribution to the acceleration.

\subsection{Metric $f(R)$ gravity}

We now consider the case of metric $f(R)$ gravity, described by the action
\begin{equation}
 S'=\frac{1}{16\pi} \int d^4x \sqrt{-g} f(R)
+S^{(matter)} \,,
\end{equation}
where $f(R)$ is a non linear function of its argument \cite{review}. It is well known that this theory is equivalent to an $\omega=0$ Brans--Dicke theory with Brans--Dicke scalar $\phi\equiv f'(R)$ and potential $V(\phi)=Rf'(R)-f(R)$~ \cite{STequivalence}. For the sake of illustration, let us take into account the Lagrangian density in the form $f(R)=R+\alpha R^n$ with $n>1$ and $\alpha>0$ as required for local stability~\cite{stability}. Then, the potential can be expressed as 
\be
V(\phi)=\frac{n-1}{n^{\frac{n}{n-1}}\alpha^{\frac{1}{n-1}}}\left(\phi-1\right)^{\frac{n}{n-1}}
\ee
and eq.~(\ref{mammamia}) reduces to
\be \label{ahiahi}
\hspace{-1cm}\frac{\ddot{a}_D}{a_D}=-\frac{16}{9}\left\langle\frac{\rho}{\phi}\right\rangle_D+\frac{\mathcal{Q}_D}{3}-\frac{1}{3}\left\langle\frac{\ddot{\phi}}{\phi}\right\rangle_D+\frac{2n-1}{18\, n^{\frac{n}{n-1}}}\, \frac{1}{\alpha^{\frac{1}{n-1}}}\left\langle(\phi-1)^{\frac{1}{n-1}}\right\rangle_D\,.
\ee
$\alpha$ arises from quantum corrections and is presumably small, so it would seem that the last term on the right hand side of the previous equation is large. However, this is not the case because $(\phi-1)^{\frac{1}{n-1}}$ is also small and contains the same power of $\alpha$: in fact, by expressing $(\phi-1)$ as a function of $R$, the last term of eq.~(\ref{ahiahi}) is rewritten as $\frac{2n-1}{18n}\left\langle R\right\rangle_D$. Nevertheless, it is relevant that this term is not suppressed by positive powers of $\alpha$, as one might expect, and hence it may contribute significantly to the cosmic acceleration. The third term on the right hand side, for small values of $\alpha$, is instead
\be
-\frac{1}{3}\left\langle\frac{\ddot{\phi}}{\phi}\right\rangle_D\simeq-\frac{1}{3}\alpha n(n-1)\left\langle(n-2)R^{n-3}\dot{R}^2+R^{n-2}\ddot{R}\right\rangle_D\,.
\ee
For the physically well-motivated case $n=2$ associated to Starobinsky inflation in the early universe~\cite{Starobinsky}, this term reduces to $-\frac{2\alpha}{3}\left\langle\ddot{R}\right\rangle_D$ and hence it is subdominant with respect to the last term of eq.~(\ref{ahiahi}). 
Finally for the first two terms on the right hand side of eq.~(\ref{ahiahi}) the same considerations presented after eq.~(\ref{mammamia}) apply.

\section{Conclusions}

The increasing improvement in quality and quantity of the cosmological data motivates a proper evaluation of the backreaction of matter inhomogeneities.
Hence, any test of alternative theories of gravitation will have to take into account possible corrections due to the backreaction mechanism,
whether the latter are large or not. For this reason, we analyzed here the possibility of improving the averaging scheme in the prototypical alternative theories of gravity, the scalar-tensor ones. 

Keeping this goal in mind and following the path outlined by Buchert and collaborators, we  have derived two
scalar  equations (the Hamiltonian constraint and the equation for the scale factor) from contractions of the  field
equations written in the form of effective  Einstein equations. 
The more general working frame exposed an intrinsic ambiguity of the averaging proposal related to the scalar degree of freedom in scalar-tensor theories.
The ambiguity is twofold as it leads to different averaged equations for different conformal frames and, within a chosen frame, to different results depending on the way the field equations are cast at the beginning of the calculation.
We made here the choice of working in the Jordan conformal frame and later on in the calculation  the \textit{ansatz} of  a domain-independent scalar field allowed us to circumvent the  ambiguity linked to the non-commutativity of the operations involved.  

As in GR, the system of equations obtained is not  closed,  
hence one extra assumption is needed in order to solve it.
The backreaction term ${\cal Q}_D$, and other terms as well, have 
signs that are undetermined and no clear effect. This is not too 
surprising, considering that a loss of information is unavoidable 
whenever an average is performed. Averaging makes it impossible 
to disentangle the individual contributions of inhomogeneties and 
anisotropies, but here even the collective effects are 
uncertain. 
While no definitive conclusion can be reached (as in the GR 
case), nonetheless there are encouraging new terms in scalar-tensor cosmology. In particular, 
we noticed that the term including the scalar field potential 
and its derivative could significantly affect the acceleration. 

In order to gain a better understanding of the potentialities of the backreaction terms in eq.~(\ref{mammamia})  to contribute significantly to late time acceleration we finally specialized to two specific sub-cases, namely Brans-Dicke and metric $f(R)$ gravity.
In the first case we have provided, as a proof of principle, a toy model solution which is accelerated at late times due to the presence of the Brans--Dicke scalar field $\phi$. 
In the second case, we have studied a polynomial Lagrangian using 
the connection between metric $f(R)$ and scalar-tensor theories.  
While it is natural to expect that higher order corrections to 
the Einstein--Hilbert Lagrangian would be suppressed by their small dimensional coefficients, we found that a generic $\alpha R^n$ term contributes via the potential term in eq.~(\ref{mammamia}) without showing any suppression in $\alpha$. Moreover, the fact that this term is now proportional to the averaged Ricci scalar implies that it is not necessarily small at late times.

The analysis outlined here would certainly benefit from  exact solutions --- even simplified toy models such as Lemaitre--Tolman--Bondi solutions \cite{LTB} --- in order to better
understand the role of matter inhomogeneities in scalar-tensor theories. The study of these exact models  will be pursued elsewhere.

\section*{Acknowledgments}

We thank a referee for useful comments. VF acknowledges the 
hospitality of SISSA and financial support from the Natural 
Sciences and  Engineering Research Council of Canada (NSERC).


\section*{References}
\begin{thebibliography}{99}

\bibitem{SN} Riess A G 1998{\em et al.} {\em Astron. J.}
{\bf 116} 1009; 1999 {\em Astron. J.} {\bf 118} 2668;
2001 {\em Astrophys. J.} {\bf 560} 49;  2004 {\em Astrophys. J.}
{\bf 607} 665;
Perlmutter S {\em et al.} 1998 {\em Nature} {\bf 391} 51;
1999 {\em Astrophys. J.} {\bf 517} 565; Tonry J L {\em et
al.} 2003 {\em Astrophys. J.} {\bf 594} 1;
Knop R {\em et al.} 2003 {\em Astrophys. J.} {\bf 598} 102;
Barris B {\em et al.} 2004 {\em Astrophys. J.} {\bf 602} 571.

\bibitem{Linderresletter} See Linder E V 2008 {\em Am. J.
Phys.} {\bf 76} 197 for a comprehensive list of references.

\bibitem{CCT} Capozziello S, Carloni S and Troisi A 2003 {\em
Recent  Res. Dev. Astron. Astrophys.} {\bf 1} 625
[arXiv:astro-ph/0303041].

\bibitem{CDTT}  Carroll S M,  Duvvuri V, Trodden M and
Turner M S 2004 {\em Phys. Rev. D} {\bf 70} 043528.

\bibitem{Buchert2}
Buchert T 2000 \grg {\bf 32} 105  [arXiv:gr-qc/9906015].

\bibitem{rasanen}  R\"{a}s\"{a}nen S 2004 {\em J. Cosmol. Astrop.
Phys.} {\bf 02} 003 [arXiv:astro-ph/0311257].

\bibitem{Buchertall} Buchert T and  Carfora M 2008
{\em Class. Quantum Grav.} {\bf 25} 195001
 [arXiv:0803.1401]; 2002 {\em Class. Quantum Grav.} {\bf 19} 6109
 [arXiv:gr-qc/0210037]; Wiltshire D L 2007 {\em New J. Phys.} {\bf
9} 377  [arXiv:gr-qc/0702082]; 2007 {\em
Phys. Rev. Lett.} {\bf 99} 251101; Kolb E W, Matarrese S, Riotto A 2006 {\em New J. Phys.} {\bf 8} 322;
 Zalaletdinov R M 1992
\grg {\bf 24} 1015; Li N and Schwarz D J 2008  {\em
Phys. Rev. D} {\bf 78} 083531;
2007 {\em Phys. Rev. D} {\bf 76} 083011; Paranjape A and Singh T P 2007
{\em Phys. Rev. D} {\bf 76} 044006.

\bibitem{Buchertexactmodels}  Marra V, Kolb E and Matarrese S
2008  {\em Phys. Rev D} {\bf 77} 023003 [arXiv:0710.5505],
Marra V, Kolb E,  Matarrese S and Riotto A 2007 {\em Phys. Rev D}
{\bf 76} 123004 [arXiv:0708.3622]; Marra V 2008 [arXiv:0803.3152];
Kolb E, Marra V and Matarrese S 2009 [arXiv:0901.4566].

\bibitem{ellis} Ellis G F R 1984 in {\em General Relativity and
Gravitation},  B Bertotti ed.  (Dordrecht: Reidel).

\bibitem{ellis2} Ellis G F R and  Stoeger W 1987 {\em Class.
Quantum Grav.} {\bf 4} 1697.

\bibitem{GreenSchwarzWitten} Green M B, Schwarz G H and Witten
E 1987 {\it Superstring Theory} (Cambridge: Cambridge  University
Press).

\bibitem{BD}  Brans C H  and  Dicke R H 1961  {\em  Phys. Rev.}
{\bf 124} 925.

\bibitem{ST}  Bergmann P G 1968 {\em Int. J. Theor. Phys.}
{\bf 1} 25; Wagoner R V 1970 {\em Phys. Rev. D} {\bf 1}
3209;  Nordvedt K 1970 {\em Astrophys. J.} {\bf 161} 1059.

\bibitem{bosonic} Callan C G,  Friedan D,   Martinez E J and
Perry M J 1985 {\em Nucl. Phys. B} {\bf 262} 593;
Fradkin E S and Tseytlin A A 1985 {\em Nucl. Phys. B} {\bf
261} 1.

\bibitem{Will} Will C M 1993, {\em Theory and Experiment in
Gravitational Physics} (Cambridge: Cambridge University Press);\\
  Ibidem,
  %``The confrontation between general relativity and experiment,''
  Living Rev.\ Rel.\  {\bf 9}, 3 (2005)
  [arXiv:gr-qc/0510072].

\bibitem{CardoneEsposito} Cardone V and Esposito G
[arXiv:0805.1203].

\bibitem{ren-group} Reuter M and Saueressig F 2007 [arXiv:0708.1317]; Litim D F 2008 [arXiv:0810.3675].

\bibitem{Starobinsky} Starobinsky A A 1980 {\em Phys. Lett. B} {\bf 91} 99
% Starobinsky A A 1981 {\em JETP Lett.} {\bf 34} 438

% 1981 {\em Sov. Astron.Lett.} {\bf 7} 36.

\bibitem{STequivalence} Higgs P W 1959 {\em Nuovo Cimento}
{\bf 11} 816;   Teyssandier P and Tourrenc P 1983
{\em J. Math. Phys.} {\bf 24}, 2793;  Whitt B 1984 {\em
Phys. Lett. B} {\bf 145} 176;   Barrow J D and
Cotsakis S 1988 {\em Phys. Lett. B} {\bf 214} 515;
Barrow J D 1988  {\em Nucl. Phys. B} {\bf 296} 697;
Wands D 1994  {\em Class. Quantum Grav.} {\bf 11} 269;
Chiba T 2003 {\em Phys. Lett. B} {\bf 575} 1.

\bibitem{review} Sotiriou T P and  Faraoni V, to appear in {\em
Rev. Mod. Phys.} [arXiv:0805.1726];\\Nojiri S and Odintsov S D,
  %``Dark energy, inflation and dark matter from modified F(R) gravity,''
  arXiv:0807.0685 [hep-th].

\bibitem{Wald} Wald R M  1984 {\it General Relativity} (Chicago:
Chicago University Press).

\bibitem{gauge} Larena J [arXiv:0902.3159]; Brown I, Behrend J
and Malik K [arXiv:0903.3264].

\bibitem{Larena08} Larena J, Alimi J-M, Buchert T, Kunz M and
Corasaniti P [arXiv:0808.1161].

\bibitem{para} Paranjape A and  Singh T P 2008 \grg {\bf 40} 139
[arXiv:astro-ph/0609481].

\bibitem{Buchert1} Buchert T 2008  \grg {\bf 40} 467
[arXiv:0707.2153]; 2007  {\em AIP Conf. Proc.} {\bf 910} 361
 [arXiv: gr-qc/0612166].

\bibitem{BuchertLarenaAlimi06} Larena J, Buchert T and Alimi
J-M 2006 {\em Class. Quantum Grav.} {\bf 23} 6379.

\bibitem{coupling} Nordtvedt K 1968 {\em Phys. Rev. D} {\bf 169}
1017.

\bibitem{BertottiIessTortora} Bertotti B,  Iess L and
Tortora P 2003 {\em  Nature} {\bf 425} 374.

\bibitem{BuchertDM} Buchert T 1996 in {\em Mapping, Measuring, 
and Modelling the Universe}, ASP Conference Series vol.~94, Coles 
P, Martinez V and Pons-Borderia M J eds. 

\bibitem{Tsagasetal08} Tsagas C G, Challinor A and Maartens R 
2008 {\em Phys. Repts.} {\bf 465} 61

\bibitem{BDlimitproblems} Faraoni V 1998 {\em Phys. Lett. A}
{\bf 245} 26; 1999 {\em Phys. Rev. D} {\bf 59} 084021.

\bibitem{stability}
  Dolgov A D and Kawasaki M,
  %``Can modified gravity explain accelerated cosmic expansion?,''
  Phys.\ Lett.\  B {\bf 573}, 1 (2003)
  [arXiv:astro-ph/0307285];\\
  Faraoni V,
  %``Matter instability in modified gravity,''
  Phys.\ Rev.\  D {\bf 74}, 104017 (2006)
  [arXiv:astro-ph/0610734];\\Nojiri S and Odintsov S D,
  %``Modified gravity with negative and positive powers of the curvature:
  %Unification of the inflation and of the cosmic acceleration,''
  Phys.\ Rev.\  D {\bf 68}, 123512 (2003)
  [arXiv:hep-th/0307288].

\bibitem{LTB} Paranjape A and Singh T P 2006 {\em Class. Quantum Grav.}
{\bf 23} 6955 [arXiv: astro-ph/0605195]; R\"{a}s\"{a}nen S 2004 {\em J. Cosmol. Astrop.
Phys.} {\bf 11} 010 [arXiv: gr-qc/0408097].


\end{thebibliography}
\end{document}